\documentclass[conference]{IEEEtran}
\ifCLASSINFOpdf
\else
\fi
\usepackage{amsmath}
\usepackage{amssymb}

\usepackage{algorithmic}
\usepackage[dvips]{graphicx}
\usepackage[latin1]{inputenc}
\usepackage{amssymb,amsmath,array}

\usepackage{amssymb}
\usepackage[english]{babel}
\usepackage{amsfonts}
\usepackage{makeidx}
\usepackage{bbm}
\usepackage{multirow}
\usepackage{caption}
\usepackage{amssymb,amsmath,amsfonts}
\usepackage[left,modulo]{lineno}
\usepackage{bm}
\usepackage{psfrag}
\usepackage{graphics}
\usepackage{fancyhdr}
\usepackage{eucal}
\usepackage[english]{babel}
\usepackage{ifpdf}
\usepackage{makeidx}
\usepackage{setspace}
\usepackage{float}
\usepackage[T1]{fontenc}
\usepackage{subfigure}
\usepackage{dsfont}
\usepackage[ruled, vlined, linesnumbered, longend]{algorithm2e}
\usepackage{textcomp}
\usepackage{booktabs}
\usepackage{multirow}
\usepackage{colortbl}

\usepackage{epstopdf}

\usepackage{tikz}
\usepackage{tikz-inet}
\usetikzlibrary{shapes,snakes,arrows,automata}

\newif\ifnotes\notestrue

\def\hgr#1{}

\newcommand{\ben}{\begin{enumerate}}
\newcommand{\een}{\end{enumerate}}

\newcommand{\bc}{\begin{center}}
\newcommand{\ec}{\end{center}}

\newcommand{\bit}{\begin{itemize}}
\newcommand{\eit}{\end{itemize}}

\newcommand{\ds}{\displaystyle}
\newcommand{\beq}{\begin{equation}}
\newcommand{\eeq}{\end{equation}}

\newcommand{\wre}{\mathbf{w}^{\rm{r}}}
\newcommand{\wi}{\mathbf{w}^{\rm{in}}}
\newcommand{\wo}{\mathbf{w}^{\rm{out}}}

\newcommand{\x}{\mathbf{x}}
\newcommand{\y}{\mathbf{y}}

\newcommand{\Nx}{d}

\newcommand{\State}{\mathbf{s}}

\newcommand{\R}{\mathds{R}}

\begin{document}
\IEEEoverridecommandlockouts
%
\title{Prediction of Facebook Post Metrics using Machine Learning\thanks{This is a draft version of a manuscript accepted in 
the XXI International Conference on Soft Computing and Measurement (SCM'2018), Saint Petersburg, Russia, May 23 - 25, 2018 (\texttt{http://scm.eltech.ru/}).
}}
%

\author{\IEEEauthorblockN{Emmanuel Sam\IEEEauthorrefmark{1},
Sergey Yarushev\IEEEauthorrefmark{2},
Sebasti\'an Basterrech\IEEEauthorrefmark{3}, 
Alexey Averkin\IEEEauthorrefmark{4}
}
\IEEEauthorblockA{\IEEEauthorrefmark{1}
Faculty of Computing and Information Technology\\
Wisconsin International University College, Accra, Ghana\\
Email: {\texttt{elsam@wiuc-ghana.edu.gh}}
}
\IEEEauthorblockA{\IEEEauthorrefmark{2}
Department of Informatics, Plekhanov Russian University of Economics\\
Moscow, Russian Federation\\
Email: {\texttt{Sergey.Yarushev@icloud.com}}
}
\IEEEauthorblockA{\IEEEauthorrefmark{3}
Department of Computer Science, Faculty of Electrical Engineering\\
Czech Technical University, Prague, Czech Republic\\
Email: {\texttt{Sebastian.Basterrech@fel.cvut.cz}}
}
\IEEEauthorblockA{\IEEEauthorrefmark{4}
Institute of Informatics Problems, FRC CSC RAS\\
Moscow, Russian Federation\\
Email: {\texttt{verkin2003@inbox.ru}}
}
}


\maketitle
\begin{abstract}
In this short paper, we evaluate the performance of three well-known Machine Learning techniques for 
predicting the impact of a post in Facebook.
Social medias have a huge influence in the social behaviour. 
Therefore to develop an automatic model for predicting the impact of posts in social medias can be useful to the society.
In this article, we analyze the efficiency for predicting the post impact of three popular techniques: Support Vector  Regression (SVR), Echo State Network (ESN) and Adaptive Network Fuzzy Inject System (ANFIS).
The evaluation was done over a public and well-known benchmark dataset.\\

\end{abstract}


%
\IEEEpeerreviewmaketitle

\section{Introduction}
Nowadays social medias impacts in the collective behaviour, and they have a very important role in the diffusion of information.
For this reason, an automatic method for predicting the impact of posts in social medias can be useful in several areas such as: marketing, phycology, educational domains, security and so on.

In this study, we analyze other Machine Learning tools for predicting a group of Facebook metrics. 
The goal is to make an automatic prediction of the impact of a post in Facebook.
A previous study in~\cite{Moro2016} explored the possibility of predicting some Facebook metrics with Support Vector Regression (SVR). 
Few of the metrics obtained good results when the evaluation is made using mean absolute percentage error.
In this study we predict $3$ metrics: \textit{Comments}, \textit{Shares}, \textit{Likes}.
These metrics are important reference of the post impact. In~\cite{Moro2016} was defined a kind of measure named \textit{Total interactions of a post} which is defined like the sum of \textit{Comments}, \textit{Shares}, \textit{Likes}.
In this short paper, we present results obtained by three popular learning methods. The Support Vector  Regression (SVR) which is based in kernels. Echo State Network (ESN) that is technique based in the power of recurrent Neural Networks and Linear Regressions. Furthermore, we evaluate Adaptive Network Fuzzy Inject System (ANFIS) that is distributed parallel system with fuzzy rules.

The rest of this paper is organized as follows. Section~\ref{ML} reviews how the machine learning techniques to be explored in this study have been applied in other studies related to social media metrics. Section~\ref{Results} provides a description of the techniques and their properties as well as the methodology for this study. Data description, result of analysis, and related discussions are presented in Section 4. Conclusions and recommendations for future work are given in Section 5.

\section{Background on Machine Learning Methods}
\label{ML} 

\subsection{Problem formalization}
\label{Formalization}
Let $\x{(t)}$ be a $p$-dimensional input data, and let $y{(t)}$ be an unidimensional output variable.
Given a learning dataset composed by $T$ real input-output pairs $(\x(t),y(t))$. The goal is to define a model $\varphi(\cdot)$ for predicting an outcome variable based on a set of input features.
Note that, we have several output variables and we are modeling each of them independently.
The model is evaluated using a quantitative measure called \textit{cost function} that measures the quality of the learning model.
In this article, we use the most popular metric when the output variable is a real value, the Mean Squared Error ($E_{\rm{MSE}}$):
\begin{equation}
E_{MSE}=\ds{\frac{1}{T}\sum_{t=1}^{T}(\hat{y}(t)-y(t))^2 },
\end{equation}
where $\hat{y}(t)$ denotes the prediction for the input $\x(t)$.
%
\subsection{Adaptive Network Fuzzy Inject System}
\label{ANFIS}
\begin{figure}[!t]
\centering
\includegraphics[width=3.3in]{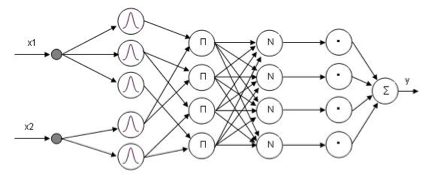}
\caption{\label{Fig1}Example of ANFIS architecture.}
\end{figure}

The Adaptive-Network-Fuzzy Inject System (ANFIS)
ANFIS is the abbreviation Adaptive-Network-Fuzzy Inject System - an adaptive network of fuzzy output. 
Proposed in the early nineties, ANFIS is one of the first variants of hybrid neural-fuzzy networks - a neural network of direct signal propagation of a special type. 
The architecture of the neural-fuzzy network is isomorphic to the fuzzy knowledge base. 
Neuro-fuzzy networks use differentiated implementations of triangular norms (multiplication and probabilistic OR), as well as smooth functions. 
This allows the use of cross-fuzzy neural networks, rapid algorithms for learning neural networks, based on the method of back propagation of errors. 
The architecture and rules for each layer of the ANFIS network are described below.
ANFIS implements the Sugeno fuzzy inference system in the form of a five-layer neural network of direct signal propagation. 
The system works as follows:
\begin{itemize}
\item the first layer is the terms of the input variables;\\
\item the second layer - antecedents (parcels) of fuzzy rules;
\item the third layer - the normalization of the degree of implementation of the rules;
\item the fourth layer is the conclusion of the rules;
\item the fifth layer is the aggregation of the result, du according to different rules.
\end{itemize}
The network inputs in a separate layer are not allocated. 
Figure~\ref{Fig1} shows an example of an ANFIS network with two input variables ($x_1$ and $x_2$) and four fuzzy rules. 
In the example, the linguistic evaluation of the input variable $x_1$, three terms are used, and for the variable $x_2$ are used two terms.

\subsection{Support Vector Regression} 
\label{SVM}
Support Vector Regression (SVR) is a version of the well-known Support Vector Machine (SVM)~\cite{Vapnik95}. The SVR model was  proposed by~\cite{Drucker1996}, it is a technique to be applied to the regression case. 
Similar to SVM, the SVR algorithm uses nonlinear mappings termed as kernels to transform an input space into a high dimensional feature space. It constructs a regression model using subset of the training instances termed as support vectors~\cite{Hastie2009}. 
The technique uses a global parameter $\varepsilon$ which is to learn a function $f(x)$ which is at most $\varepsilon$ deviations away from the target by defining a band around the regression function. 
Another global parameter, denoted by $C$, controls the trade-off between the prediction error and the flatness of the band around $f(\x(t))$. 
Finally a test instance $\x(t)$ can be predicted using the following equation:
\begin{equation}
f(\x(t)) = \beta_0 + \sum\limits_i^q {\beta_i K(\x(i) ,\x(t))}, \text{ for } \quad i=1,2...q,
\end{equation}
where: $x_i$ are the support vectors (points that fall outside or on the border of the tube), $\x(t)$ is a test instance to be predicted, $\mathbf{\beta}$ is the vector of parameters determined by the SVR learning algorithm, and $K(\cdot ,\cdot)$ is a kernel function used to transform the input data points into a higher dimensional feature space~\cite{Vapnik95}.

SVM has been applied in social network analysis for the classification of Chinese Facebook users into introverts and extroverts based on their Facebook wall posts~\cite{Peng2015}. SVM has also outperformed other classifies in several comparative analysis in the context of social media~\cite{Joshi2016}. SVR is has been applied in~\cite{Moro2016} for the  prediction of social media performance metrics~\cite{Moro2016}.

\subsection{Echo State Network}
\label{ESQN}
Since the early 2000s, Reservoir Computing (RC) has gained prominence in the Neural Network community~\cite{Jaeger09}.
A RC model has a dynamical system called \textit{reservoir}, which expands input data into a high-dimensional space in a similar way that kernels methods. 
%
%
%
Next, the model uses a supervised learning tool to predict the model outputs. 
Most often, the RC model uses a simple linear regression from the feature map and the output space.
RC models have been widely used in fields such as:  pattern classification~\cite{Jaeger01}, speech recognition~\cite{Verstraeten07, Maass03}, speech quality~\cite{Baster13PESQ} and time-series prediction~\cite{Jaeger09,Jaeger01,Steil04,Schrauwen07,BasterCord11}, the Internet traffic prediction~\cite{Baster12ESQN}, and so on.
%
%
%
%
%
Most formal, a reservoir is a temporal expansion function from an input space $\R^{p}$ into a larger space $\R^{d}$ with~$p\ll d$. 
We denote by $\x(t)\in\R^{p}$ the input pattern of the model at any time $t$ and $\y(t)$ the target of the model at any time $t$. 
The recurrences are modelled with a state vector $\State(t)\in\R^{d}$:
\begin{equation}
\label{reservoirState}
\State{(t)}=\phi(\State{(t-1)},\x{(t)}),
\end{equation}
where $\phi(\cdot)$ is an expansion function.

\newcommand{\wrepmi} {{w^{\rm{r+}}_{mi}}}
\newcommand{\wrenmi} {{w^{\rm{r-}}_{mi}}}
\newcommand{\wipmi} {{w^{\rm{in+}}_{mi}}}
\newcommand{\winmi} {{w^{\rm{in-}}_{mi}}}
\newcommand{\wrepji} {{w^{\rm{r+}}_{ji}}}
\newcommand{\wrenji} {{w^{\rm{r-}}_{ji}}}
\newcommand{\wipji} {{w^{\rm{in+}}_{ji}}}
\newcommand{\winji} {{w^{\rm{in-}}_{ji}}}
\newcommand{\wrep}{{\mathbf{w}^{\rm{r}+}}}
\newcommand{\wren}{{\mathbf{w}^{\rm{r}-}}}
\newcommand{\wip}{{\mathbf{w}^{\rm{in}+}}}
\newcommand{\win}{{\mathbf{w}^{\rm{in}-}}}
%
%
%
We denote the model connections by ${\wi}$ and  ${\wre}$  which are matrices of dimensions $d\times p$ (for the input weights) and $d \times d$ (for the reservoir weights).
A characteristic of the model is that these weight matrices are fixed during the training algorithm~\cite{Jaeger01}. They are randomly initialized and they kept fixed.
%
%
%
To compute the ESN output corresponding to a new input $\x(t)$ (a column vector) , the model first computes a new reservoir state $\mathbf{s}(t)=(s_1(t),\ldots,s_{\Nx}(t))^t$ computed by
\begin{equation}
\label{ESNreservoirState}
\mathbf{s}(t)=\tanh(\wre \mathbf{s}(t-1) + \mathbf{w^{\rm{in}}}\x(t)).
\end{equation}
%
%
The vector $\wo$ is the only adjustable parameters in ESN model, which is usually estimated using ridge regression between the vector $[1,\mathbf{s}]$ and the target.
%
%


\section{Results of Experiments}
\label{Results}
\subsection{Description of Data}
The dataset contains~$7$ features known prior to post publication, and~$3$ output variables which are used for the post impact. 
The output variables are: \textit{comments}, \textit{shares}, and \textit{likes}.
The variable \textit{comments} counts the number of comments that provoked a specific post.
The variable \textit{shares} refers to the number of times that the post has been shared with other users.
The variable \textit{likes} is also a counter operation, that counts the number of likes caused by a post.
\begin{table}[h]
\begin{center}
	\tiny
	\caption{\label{data}Facebook post performance metrics.}
	\begin{tabular}{ccccccc}
		\toprule
		{Variable} & Mean & Median & Mode & St. deviation & Maximum & Minimum \\
		\midrule
		{Number of comments} & \multicolumn{1}{r}{7} & \multicolumn{1}{r}{3} & \multicolumn{1}{r}{0} & \multicolumn{1}{r}{21} & \multicolumn{1}{r}{372} & \multicolumn{1}{r}{0} \\
		\midrule
		{Number of likes} & \multicolumn{1}{r}{178} & \multicolumn{1}{r}{101} & \multicolumn{1}{r}{98} & \multicolumn{1}{r}{323} & \multicolumn{1}{r}{5172} & \multicolumn{1}{r}{0} \\
		\midrule
		{Number of shares} & \multicolumn{1}{r}{27} & \multicolumn{1}{r}{19} & \multicolumn{1}{r}{13} & \multicolumn{1}{r}{43} & \multicolumn{1}{r}{790} & \multicolumn{1}{r}{0} \\
		\bottomrule
	\end{tabular}
\end{center}
\end{table}

\subsection{Discussion of results}
\label{Discussions}
%
The training of ESN was done with a reservoir size of 25 neurons and spectral radius of $0.5$ (both parameters are important in the design of the model~\cite{Jaeger09}).
A similarly number of hidden neuron has the ELM model.
The SVR algorithm applied in this study uses a Gaussian radial basis (RBF) kernel function. The width of the Gaussian kernel $\gamma$ was set to 0.1. The band defined around the regression function, $\epsilon$, was set to 0.1, and $C$ the parameter which controls the trade-off between the prediction error and the flatness of the band around the regression function was set to 1000. The selection of these values for the parameters was based on~\cite{scikit-learn}.
The training of ANFIS was done with 400 data pairs. ANFIS has 7 input layers, 1 output layer. In input Member Function type was selected the Gauss method, and we take 3 member functions for each input layer. 
In the output layer we are selected constant MF type. 
For training network we are using a hybrid optimization method. 
The training of the model parameters required 2 epoch of the training dataset.

Table~\ref{Results} presents the obtained MSE of the three learning techniques applied in this study.
The new propose techniques in this article (ESN and ANFIS) obtain better results than SVR. Although, 
ANFIS seems to performs better for predicting the amount of \textit{likes}, the ESN model has a better accuracy in the other two cases.
Figure~\ref{Results} present an illustration of our results, where is easy to compare the different accuracy among the models.
\begin{table}[h]
\begin{center}
	\caption{\label{Results}Prediction of the Machine Learning tools.}
	\begin{tabular}{cccc}
		\toprule
		{Method} & Comments & Likes & Shares\\
		\midrule
		{SVR} & 0.002998 & 0.003917 & 0.003496 \\
		\midrule
		{ESN} & \textbf{0.002068} & 0.003163 & \textbf{0.001740} \\
		\midrule
		ANFIS    & 0.0022092   & \textbf{0.002121} & 0.001957\\
		\bottomrule
	\end{tabular}
\end{center}
\end{table}

\begin{figure}[!t]
\centering
\includegraphics[width=3in]{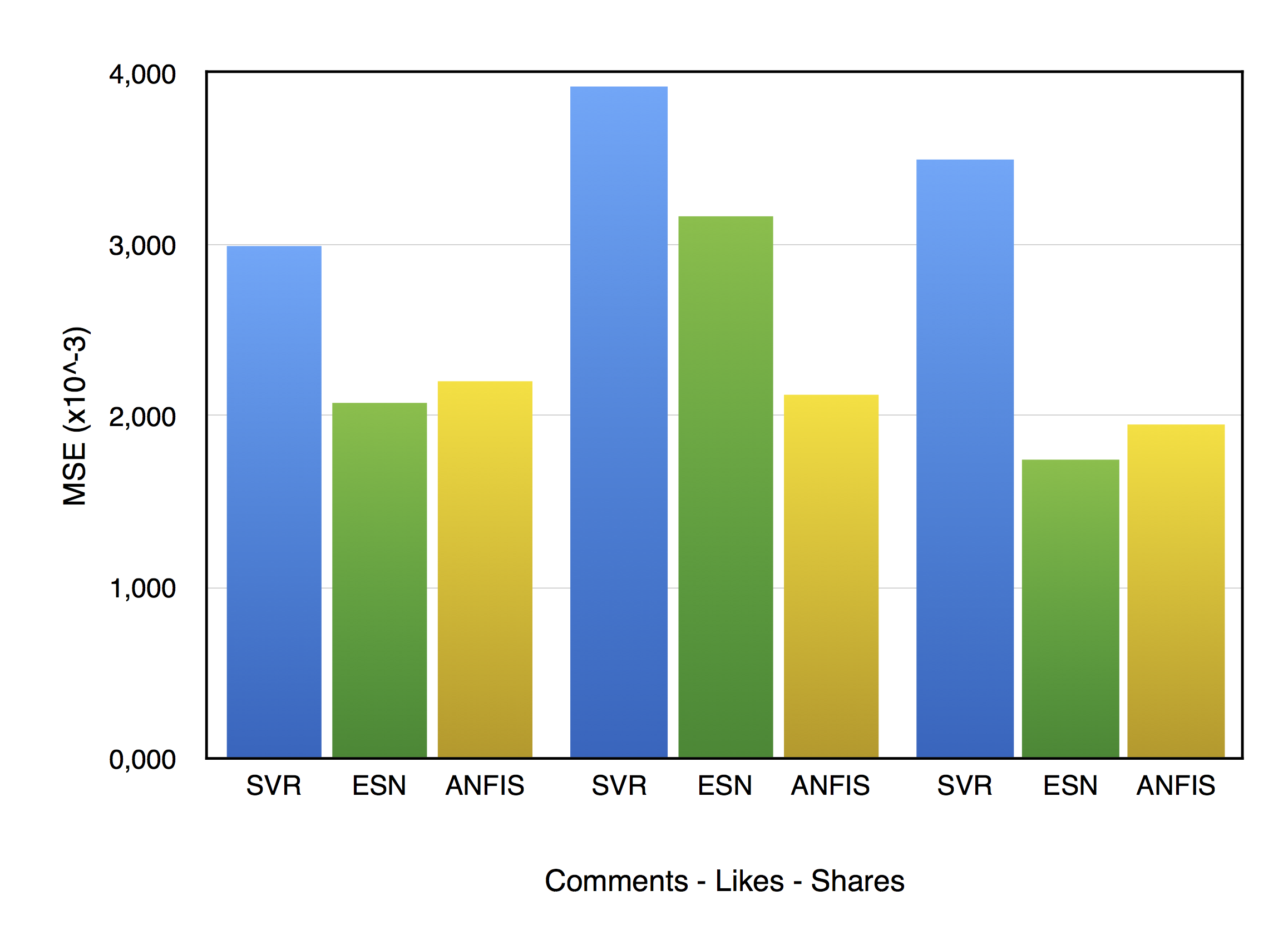}
\caption{\label{Results}Accuracy of the three learning tools for predicting the output variables.}
\end{figure}

\section{Conclusions and Future Works}
\label{Conclusions}
In this work, we are showed prediction accuracy of three models - Support Vector Regression, Echo State Network and Adaptive Neuro-Fuzzy Inference System. For example we have predicted the impact of a post in social network Facebook. The dataset contains 7 features known prior to post publication, and 3 output variables which are used for the post impact. The output variables are: comments, shares, and likes. The new propose techniques in this article (ESN and ANFIS) obtain better results than SVR. Although, ANFIS seems to performs better for predicting the amount of likes, the ESN model has a better accuracy in the other two cases. In future work we will to continue experiments with using ANFIS model for other datasets to comparise it with other second techniques. We also plan to test in this task deep neural networks.


\bibliographystyle{IEEEtran}
\bibliography{references}
%
%
%

\end{document}